\documentclass[twocolumn,prb,floatfix]{revtex4}

\usepackage{graphicx}
\usepackage{psfrag}
\usepackage{textcomp}
\usepackage{latexsym}

\begin{document}

\title{Pulse and hold strategy for switching current measurements}
\date{\today}
\author{Jochen Walter}
\author{Erik Thol\'en}
\author{David B. Haviland}
\affiliation{Nanostructure Physics, Royal Institute of Technology, AlbaNova University Center, SE-10691 Stockholm, 
Sweden}

\author{Joachim Sj\"ostrand}
\affiliation{Department of Physics, Stockholm University, AlbaNova University Center, SE-10691 Stockholm, Sweden}

\begin{abstract}
  We investigate by theory and experiment, the Josephson junction switching current detector in an environment with 
frequency dependent damping. Analysis of the circuit's phase space show that a favorable topology for switching can be 
obtained with overdamped dynamics at high frequencies.  A pulse-and-hold method is described, where a fast switch 
pulse brings the circuit close to an unstable point in the phase space when biased at the hold level.  Experiments are 
performed on Cooper pair transistors and Quantronium circuits, which are overdamped at high frequencies with an 
on-chip RC shunt.  For 20~\textmu s switch pulses the switching process is well described by thermal equilibrium 
escape, based on a generalization of Kramers  formula to the case of frequency dependent damping.  A capacitor bias 
method is used to create very rapid, 25 ns switch pulses, where it is observed that the switching process is not 
governed by thermal equilibrium noise.
\end{abstract}

\maketitle

\section{Introduction}
A classical non-linear dynamical system, when driven to a point of instability, will undergo a bifurcation, where the 
system evolves toward distinctly different final states.  At bifurcation the system becomes very sensitive and the 
smallest fluctuation can determine the evolution of a massive system with huge potential energy.  This property of 
infinite sensitivity at the point of instability can be used to amplify very weak signals, and has recently been the 
focus of investigation in the design of quantum detectors to readout the state of quantum bits (qubits) built from 
Josephson junction (JJ) circuits.  Here we examine in experiment and theory a pulse and hold strategy for rapid 
switching of a JJ circuit which is quickly brought near a point of instability, pointing out several important 
properties for an ideal detector.  We focus on switching in a circuit with overdamped phase dynamics at high 
frequencies, and underdamped at low frequencies.   This HF-overdamped case is relevant to experiments on small 
capacitance JJs biased with typical measurement leads.  

Classical JJs have strongly non-linear electrodynamics and they have served as a model system in non-linear physics 
for the last 40 years.  More recently it has been shown that JJ circuits with small capacitance can also exhibit 
quantum dynamics when properly measured at low enough temperatures.  Experimental demonstration of the macroscopic 
quantum dynamics in these circuits has relied on efficient quantum measurement strategies, characterized by high 
single shot sensitivity with rapid reset time and low back action. Some of these measurement or detection methods are 
based on the switching of a JJ circuit from the zero voltage state to a finite voltage state. 
\cite{vion:quantronium:02,chiorescu:fluxqubit:03,martinis:qubit:02,plourde:fluxqubit:05}
Other detection methods are based on a dispersive technique, where a high frequency signal probes the phase dynamics 
of a 
qubit.\cite{ilichev:continuous:03,born:spectroscopy:04,siddiqi:bifurcation:04,grajcar:fluxqubti:04,wallraff:cavity:04,roschier:phasedetector:05,duty:capcitance:05,lupascu:dispersive:
06}
These dispersive methods have achieved the desired sensitivity at considerably higher speeds than the static switching 
methods, allowing individual quantum measurements to be made with much higher duty cycle.  In particular, the 
dispersive methods have shown that it is possible to continuously monitor the qubit.\cite{wallraff:visibility:05}  
However, for both static switching and dispersive methods, the sensitivity of the technique is improved by exploiting 
the non-linear properties of the readout circuit in a pulse and hold measurement strategy.  This improved sensitivity 
of the pulse and hold method is not surprising, because when properly designed, the pulse and hold technique will 
exploit the infinite sensitivity of a non-linear system at the point of instability.  

The general idea of exploiting the infinite sensitivity at an instable point is a recurrent theme in applications of 
non-linear dynamics. The basic idea has been
used since the early days of microwave engineering in the well-known parametric 
amplifier\cite{louisell:electronics:60} which has infinite gain at the point of dynamical instability.  The unstable 
point can be conveniently represented as a saddle point for the phase space trajectories of the non-linear dynamical 
system.  In the pulse and hold measurement method an initial fast pulse is used to quickly bring the system to the 
saddle point for a particular hold bias level.  The hold level is chosen so that the phase space topology favors a 
rapid separation in to the two basins of attraction in the phase space.  The initial pulse should be not so fast that 
it will cause excessive back action on the qubit, but not so slow that it's duration exceeds the relaxation time of 
the qubit.  The length of the hold pulse is that which is required to achieve a signal to noise ratio necessary for 
unambiguous determination of the resulting basin of attraction.  In practice this length is set by the filters and 
amplifiers in the second stage of the quantum measurement system.  

In this paper we discuss pulse and hold detection in the context of switching from the zero voltage state to the 
finite voltage state of a JJ.  We give an overview of such switching in JJs,  focusing on the HF-overdamped case.  
Switching detectors with overdamped high frequency phase dynamics are different from all other qubit measurement 
strategies implemented thus far, where underdamped phase dynamics has been used.  However, the HF-overdamped case is 
quite relevant to a large number of experiments which measure switching in low capacitance JJs with small critical 
currents.\cite{joyez:josephsoneffect:99,agren:prb:02,cottet:CPTsingleshot:02,mannik:periodicity:04} We show that by 
making the damping at high frequencies large enough, a favorable phase space topology for switching can be 
achieved.\cite{sjostrand:phasespace:06} In this overdamped situation the external phase can be treated classically and contributions of macroscopic quantum tunneling (MQT) to the switching probability can be neglected, in contrast to the underdamped case.\cite{ithier:zener:05}  Experimental results are shown where on-chip RC damping circuits are used to 
create an HF-overdamped environment.  We observe that for longer pulses of duration 20 \textmu s, the switching 
process is initiated by thermal fluctuations in the overdamped system and thermal equilibrium is achieved at the base 
temperature of the cryostat (25 mK).  For short pulses of duration $<25$~ns, the switching is unaffected by thermal 
fluctuations up to a temperature of 500 mK, and the width of the switching distribution at low temperatures is rather 
determined by random variations in the repeated switch pulse.  Although the detector apparently had the speed and 
sensitivity required for making a quantum measurement, we were unfortunately unable to demonstrate quantum dynamics of 
the qubit due to problems with fluctuating background charges.

\section{Phase Space Portraits}
\begin{figure}
\begin{center}
\includegraphics[width=0.35\textwidth]{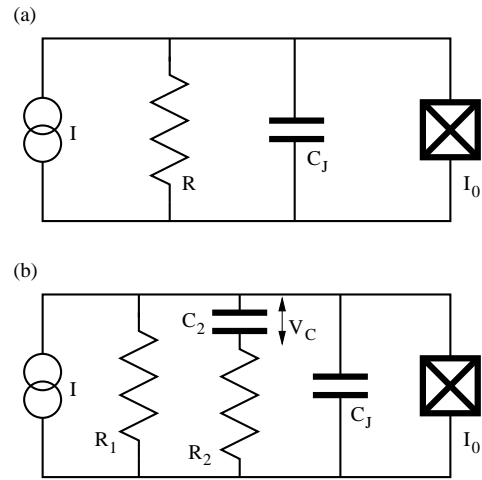}
\caption{(a) RCSJ model of a Josephson junction. (b) Simple model of a junction embedded in an environment with 
frequency dependent impedance.}
\label{washboard}
\end{center}
\end{figure}

The non-linear dynamics of a DC-driven JJ can be pictorially represented in a phase space portrait.  We begin by 
examining the phase space portraits of the resistive and capacitively shunted junction (RCSJ), which is the simplest 
model from which we can gain intuitive understanding of the non-linear dynamics.  The RCSJ model consists of an ideal 
Josephson element of critical current $I_0$ biased at the current level $I$, which is shunted by the parallel plate 
capacitance of the tunnel junction, $C_J$ and a resistor $R$, which models the damping at all frequencies (see 
fig.~\ref{washboard}(a)). The circuit parameters define the two quantities $\omega_p=\sqrt{I_0/\varphi_0C_J}$ called 
the plasma frequency, and the quality factor $Q=\omega_pRC$.  The dynamics is classified as overdamped or underdamped 
for $Q<1$ or $Q>1$, respectively. Here $\varphi_0=\hbar/2e$ is the reduced flux quantum.  

\begin{figure}[!hbp]
\begin{center}
\includegraphics[width=0.45\textwidth]{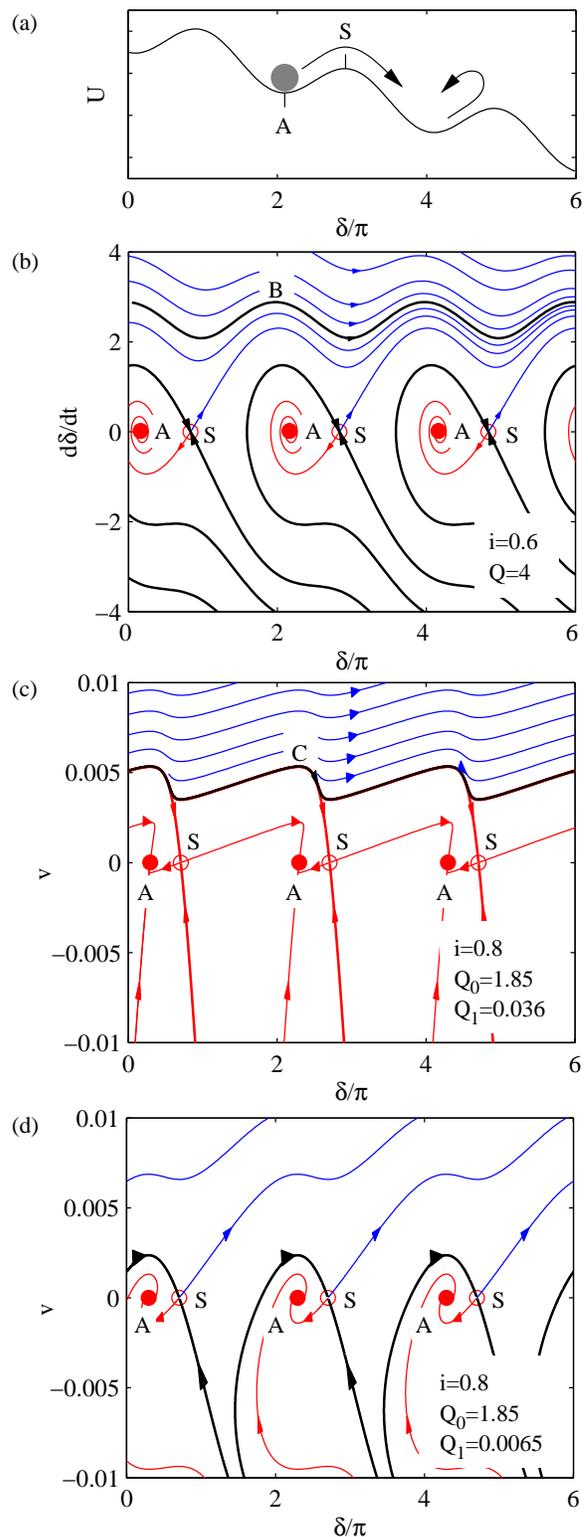}
\caption{(Color online) (a) Tilted washboard model. (b) Underdamped circuit biased at $i_r<i<1$. (c) Circuit with 
frequency dependent damping and $Q_1>Q_{1c}$ and with $Q_1<Q_{1c}$ (d).}
\label{phaseSpace1}
\end{center}
\end{figure}

The circuit dynamics can be visualized by the motion of a particle of mass $\varphi_0^2C$ in a tilted washboard 
potential $U(\delta)=-E_J(i\delta +\cos\delta)$ subjected to the damping force $\varphi_0^2/R$, where the particle 
position corresponds to the phase difference $\delta$ across the junction and the tilt $i$ is the applied current 
normalized by the critical current, $i=I/I_0$ (see fig~\ref{phaseSpace1}(a)).  Below the critical tilt $i=1$ the 
fictitious particle will stay in a local minimum of the washboard potential (marked A in fig~\ref{phaseSpace1}(a)) 
corresponding to the superconducting state where $V= \langle \dot{\delta} \rangle =0$.  Increasing the tilt of the 
potential to $i>1$, where local minima no longer exist,  the particle will start to accelerate to a finite velocity 
$V= \langle \dot{\delta}\rangle >0$ determined by the damping.   If the tilt is then decreased below $i=1$, the 
particle for an underdamped JJ ($Q>1$) will keep on moving due to inertia. Further decreasing the tilt below the level 
$i<i_r$, where loss per cycle from damping exceeds gain due to inertia, the particle will be retrapped in a local 
minimum.  In terms of the current-voltage characteristic, this corresponds to hysteresis, or a coexistence of two 
stable states, $V=0$ and $V>0$  for a bias fixed in the region $i_r<i<1$. For the overdamped RCSJ model the particle 
will always be trapped in a local minimum for $i\leq 1$ and freely evolving down the potential for $i>1$ and there is 
no coexistence of two stable states.

A phase space portrait\cite{kautz:escape:88,hanggi:review:90,sjostrand:phasespace:06} of the RCSJ model is shown in 
fig.~\ref{phaseSpace1}(b). This portrait shows trajectories that the particle would follow in the space of coordinate 
($\delta$) versus velocity ($\dot{\delta}$) for a few chosen initial conditions. The topology of the phase space 
portrait is characterized by several distinct features.  Fix point attractors marked ``A'' in fig.~\ref{phaseSpace1} 
correspond to the particle resting in a local minimum of the washboard potential, and the saddle points marked ``S'' 
corresponds to the particle resting in an unstable state at the top of the potential barrier  (compare with 
fig.~\ref{phaseSpace1}(a)).  Two trajectories surrounding A and ending at S are the unstable trajectories which define 
the boundary of a basin of attraction: All initial conditions within this boundary will follow a trajectory leading to 
A.  We call this the 0-basin of attraction.  The thick line B is a stable limiting cycle, corresponding to a 
free-running state of the phase $\delta$, where the circuit is undergoing Josephson oscillations with frequency 
$\omega = V / \varphi_0$.  All trajectories leading to the limiting cycle B start in the 1-basin of attraction, which 
is the region outside the 0-basins.  The existence of two basins of attraction in the phase space topology, and in 
particular the clear separation of all 0-basins by the 1-basin, make the underdamped RCSJ circuit ($Q>1$) appropriate 
for a switching current detector, as we discuss below.  For the overdamped RCSJ circuit ($Q<1$) attractors A and B do 
not coexist for any fixed bias condition, and it therefore can not be used for a switching current detector.  However, 
the RCSJ model is not always the most realistic model for the dynamics of JJ circuits, as the damping in real 
experiments is usually frequency dependent, and in the case of small capacitance JJs, this frequency dependence can 
very much change the character of the damping.  

At high frequencies of the order of the plasma frequency of the junction (20-100 GHz for Al/AlOx/Al tunnel junctions) 
losses are typically due to radiation phenomena, where the leads to the junction act as a wave guide for the microwave 
radiation.  If we model the leads as a transmission line, the high frequency impedance would correspond to a damping 
resistance of the order of free space impedance $Z\approx Z_0/2\pi=60~\Omega$.  With the small capacitance of a 
typical JJ as used in present experiments, this damping inevitably leads to overdamped dynamics $Q<1$.  It should be 
noted that for small capacitance JJs, underdamped phase dynamics is hard to achieve in practice as high impedance all 
the way up to the plasma frequency is desired, and this requires an engineering effort where the high impedance leads 
need to be constructed very close to the junction. \cite{corlevi:duality:06}  However, at lower frequencies (typically 
below $\approx 10$~MHz) the junction will see an impedance corresponding to the bias resistor $R$ at the top of the 
cryostat, which can be chosen large enough to give $Q>1$.  The simplest circuit which captures the frequency 
dependence described above, is a JJ shunted by a series combination of a resistor $R_2$ and a capacitor $C_2$ in 
parallel with the resistor $R_1$ as shown in fig.~\ref{washboard}(b).  At high frequencies where $C_2$ is essentially 
a short, the circuit is described by the high-frequency quality factor $Q_1=\omega_pR_{||}C_J$, where $R_{||}$ is the 
parallel combination of $R_1$ and $R_2$.
At low frequencies where $C_2$ effectively blocks, the quality factor reads $Q_0=\omega_pR_1C_J$.  This model has been 
studied previously by several 
authors.\cite{ono:currentvoltage:87,kautz:noise:90,vion:activation:96,joyez:josephsoneffect:99,sjostrand:phasespace:06
}   Casting such a circuit in a more mathematical language, it can be described by the coupled differential equations 
\cite{kautz:noise:90,sjostrand:phasespace:06}
\begin{eqnarray}
\dot{\delta} &=& \frac{Q_1}{Q_0}\left[ -\frac{1}{E_J}\frac{dU(\delta)}{d\delta}+v\left( 
\frac{Q_0}{Q_1}-1\right)+i_{n1}+i_{n2}\right]\label{diffeqn}\\
\dot{v} &=& \frac{\rho Q_1}{Q_0^3}\left[-\frac{1}{E_J}\frac{dU(\delta)}{d\delta} - v + i_{n1} + 
i_{n2}\frac{Q_1^2}{Q_0(Q_0-Q_1)}  \right],\nonumber
\end{eqnarray}
where $v=V_C/R_1I_0$ is the reduced voltage across $C_2$ and $\rho=R_1C_J/R_2C_2$ reflects the value of the transition 
frequency, being $\omega\approx 1/R_2C_2$,  between high- and low-impedance regimes. 

Phase space portraits for such a circuit are shown in figs.~\ref{phaseSpace1}(c) and (d). Here the y-axis shows the 
voltage $v$ which is directly related to $\dot{\delta}$. The topology of this phase portrait is also characterized by 
the coexistence of fix-points A and the limiting cycle B (not shown). However, for the parameters of 
fig.~\ref{phaseSpace1}(c) ($Q_0=1.85$,  $Q_1=0.036$ and $i=0.8$. ), the 0-basins and the 1-basin are now separated by 
an unstable limiting cycle C which does not intersect a saddle point.  An initial condition which is infinitesimally 
below or above C will eventually end up either in an attractor A, or on B respectively. In fig.~\ref{phaseSpace1}(c) 
we also see that the boundaries of the 0-basins are directly touching one another as a consequence of the existence of 
C. Thus, it is possible to have a trajectory from one 0-basin to another 0-basin, without crossing the 1- basin. 

This same HF-overdamped model can however produce a new topology by simply lowering the high-frequency quality factor 
$Q_1$.  As we increase the high frequency damping, the unstable limiting cycle C slowly approaches the saddle points 
S. For a critical value of $Q_1=Q_{1c}$, C and S will touch and the phase-portrait suddenly changes its topology.  
Fig.~\ref{phaseSpace1}(d) shows the phase space portrait for  $Q_0=1.85$,  $Q_1=0.0065<Q_{1c}$ and $i=0.8$ where we 
can see that C disappears and adjacent 0-state basins are again separated by the 1-basin -- a topology of the same 
form as the underdamped RCSJ model.

\section{The Switching Current Detector}

The transition from the 0-basin to the 1-basin, called switching, can be used as a very sensitive detector.  The idea 
here is to choose a "hold" bias level and circuit parameters where the phase space portrait has a favorable topology 
such as that shown in figs.~\ref{phaseSpace1}(b) and (d).  A rapid "switch pulse" is applied to the circuit bringing 
the system from A to a point as close as possible to the unstable point S.  Balanced at this unstable point, the 
circuit will be very sensitive to any external noise, or to the state of a qubit coupled to the circuit.  The qubit 
state at the end of the switch pulse can be thought of as determining the initial condition, placing the fictitious 
phase particle on either side of the basin boundary, from which the particle will evolve to the respective attractor.  
The speed and accuracy of the measurement will depend on how rapidly the particle evolves away from the unstable point 
S, far enough in to the 0-basin or 1-basin such that external noise can not drive the system to the other basin.  From 
this discussion it is clear that a phase space portrait with the topology shown in fig.~\ref{phaseSpace1}(c) is not 
favorable for a switching current detector.  Here the switching corresponds to crossing the unstable cycle C.  Initial 
conditions which are infinitesimally close to C will remain close to C over many cycles of the phase, and thus a small 
amount of noise can kick the system back and forth between the 0-basins and the 1-basin, leading to a longer 
measurement time and increased number of errors.  

The {\it measurement time} is that time which is required for the actual switching process to occur and must be 
shorter than the relaxation time of the qubit.  In the ideal case the measurement time would be the same as the 
duration of the switch pulse.  A much longer time may be required to actually determine which basin the system has 
chosen.  This longer {\it detection time} is the duration of the hold level needed to reach a signal to noise ratio 
larger than 1, which is in practice determine by the bandwidth of the low-noise amplifier and filters in the second 
stage of the circuit.  In our experiments described in the following sections, we used a low noise amplifier mounted 
at the top of the cryostat which has a very limited bandwidth and high input impedance.  While this amplifier has very 
low back action on the qubit circuit (very low current noise), it's low bandwidth increases the detection time such 
that individual measurements can be acquired only at $<10$~kHz repetition rate.  Since many measurements ($10^4$) are 
required to get good statistics when measuring probabilities, the acquisition time window is some 0.5 seconds and the 
low frequency noise (drift or $1/f$ noise) in the biasing circuit will thus play a role in the detector accuracy. 

\section{Fluctuations}\label{fluctuations}

The measurement time of a switching detector will depend on fluctuations or noise in the circuit.  The phase space 
portraits display the dissipative trajectories of a dynamical system, but they do not contain any information about 
the fluctuations which necessarily accompany dissipation.   For a switching current detector, we desire that these 
fluctuations be as small as possible, and therefore the dissipative elements should be kept at as low a temperature as 
possible.  Analyzing the switching current detector circuit with a thermal equilibrium model, we can  calculate the 
rate of escape from the attractor A.  This equilibrium escape rate however only sets an upper limit on the measurement 
time.  When we apply the switch pulse, the goal is to bring the circuit out of equilibrium, and we desire that the 
sensitivity at the unstable point be large enough so that the measurement is made before equilibrium is achieved (i.e. 
before thermal fluctuations drive the switching process).  

Equilibrium fluctuations can cause a JJ circuit to jump out of it's basin of attraction in a process know as thermal 
escape.  The random force which gives rise to the escape trajectory will most likely take the system through the 
saddle point S, because such a trajectory would require a minimum of energy from the noise source. 
\cite{kautz:escape:88}  For the topology of phase space portraits shown in figs.~\ref{phaseSpace1}(b) and (d), thermal 
escape will result in a switching from a 0-basin to the 1-basin, with negligible probability of a "retrapping" event 
bringing the system back from the 1-basin to a 0-basin.  However, for the topology of fig.~\ref{phaseSpace1}(c), 
thermal escape through the saddle point leads to another 0-basin, and thus the particle is immediately retrapped in 
the next minimum of the washboard potential.  This process of successive escape and retrapping is know as phase 
diffusion, and it's signature is a non-zero DC voltage across the JJ circuit when biased below the critical current, 
$i < 1$.  

Phase diffusion can occur in the overdamped RCSJ model, or in the HF-overdamped model when parameters result in a 
phase space topology of fig.~\ref{phaseSpace1}(c).  In the latter case, a switching process can be identified which 
corresponds to the escape from a phase diffusive state to the free running state, or to crossing the unstable limiting 
cycle C in fig.~\ref{phaseSpace1}(c) which marks the boundary between the phase diffusive region and the 1-basin.  
This basin boundary C is formed by the convergence of many trajectories leading to different S, and the escape process 
of crossing this boundary is fundamentally different than escape from a 0-basin to the 1-basin.  Numerical simulations 
\cite{sjostrand:naples:05,sjostrand:phasespace:06,sjostrand:phd:06} of switching in JJs with such a phase space 
topology show that escape over the unstable boundary C is characterized by late switching events, which arise because 
even a small amount of noise near this boundary can kick the system back and forth between the 1-basin and the many 
0-basins for a long time before there is an actual escape leading to the limiting cycle B.    
 
The rate of thermal escape from a 0-basin can be calculated using Kramers' formula 
\cite{kramers:escape:40,hanggi:review:90,melnikov:review:91} 
\begin{equation}
\Gamma=\kappa \frac{\omega_0}{2\pi}\exp (-\Delta E/k_B T),\label{Kramers}
\end{equation}
with $\Delta E$ being the depth of the potential well from A to S, $k_B$ Boltzmann's constant and $T$ the temperature. 
The prefactor  $\kappa \omega_0/2\pi$ is called the attempt frequency, where $\kappa<1$ is a factor which depends on 
the damping.  Analytical results for $\kappa$ were found by Kramers in the two limiting cases of underdamped ($Q>1$) 
and overdamped ($Q<1$) dynamics.  For the application of Kramers' escape theory we require that $\Delta E\gg k_B T$, 
i.e. thermal escape is rare, so that each escape event is from a thermal equilibrium situation.  The fluctuations in 
thermal equilibrium are completely uncorrelated in time, which is to say that the strength of the fluctuations are 
frequency independent (white noise).  Furthermore, the Kramers formula assumes absorbing boundary conditions, where 
the escape process which leads to a change of the basin of attraction has zero probability of return.  These 
conditions restrict the direct application of Kramers formula in describing switching in JJ 
circuits\cite{buttiker:activation:83} to the case of the underdamped RCSJ model such as that depicted in 
fig.~\ref{phaseSpace1}(b).  In principle one could apply Kramers formula to the overdamped RCSJ model, where the 
escape is from one well to the next well (switching between adjacent attractors A), but experiments thus far are 
unable to measure a single $2 \pi$ jump of the phase, as this corresponds to an extremely small change in circuit 
energy.  

Thermal induced switching of small capacitance Josephson junctions which experience frequency dependent damping as 
modeled by the circuit of Fig.~\ref{washboard}(b), was analyzed in experiment and theory by the Quantronics group 
\cite{vion:activation:96,joyez:josephsoneffect:99} who generalized Kramers result.  The theoretical analysis was 
subject to the constraint that the dynamics of the voltage across the shunt capacitor $v$ is underdamped (i.e. the 
quality factor $\alpha=R_2R_{||}C_2I_0/\varphi_0(R_1+R_2)\gg 1$ where $R_{||}$ is the parallel resistance of $R_1$ and 
$R_2$) so that the dynamics of $v$ is subject to the fast-time average effects of the fluctuating phase $\delta$.  
Separating timescales in this way, the switching of $v$ could then be regarded as an escape out of a meta-potential, 
$B$, formed by the averaged fluctuating force in the tilted washboard potential $F= i - \langle \sin \delta \rangle - 
v$.  Assuming non-absorbing boundary conditions, this "escape over a dissipation barrier" can be written as a 
generalization of Kramers' formula 
\begin{equation}
\Gamma=\frac{D(v_t)}{2\pi}\sqrt{\left( \frac{-F}{\lambda D}\right)'_{v_b} \left( \frac{F}{\lambda D}\right)'_{v_t} 
}\exp(B)
\label{overdampedEscape}.
\end{equation}
Here $D(v)$ is the position-dependent diffusion constant, and $B=\int_{v_b}^{v_t}(F/\lambda D)dv$, where $v_b$ and 
$v_t$ stand for the bottom and the top of the effective barrier, respectively. Detailed expressions can be found in 
refs.\cite{gotz:phd:97,joyez:josephsoneffect:99}

In section~\ref{analysis}, we use these escape rate formulas to analyze pulse and hold switching measurements.  We 
demonstrate that long switching pulses lead to thermal equilibrium switching, whereas short pulses switch the circuit 
in a way that is independent of temperature at low temperatures, with the switching distribution determined by noise 
in the switch pulse rather than noise from the cooled damping circuit. 

\section{Experiments}

\begin{figure}
\begin{center}
\includegraphics[width=0.45\textwidth]{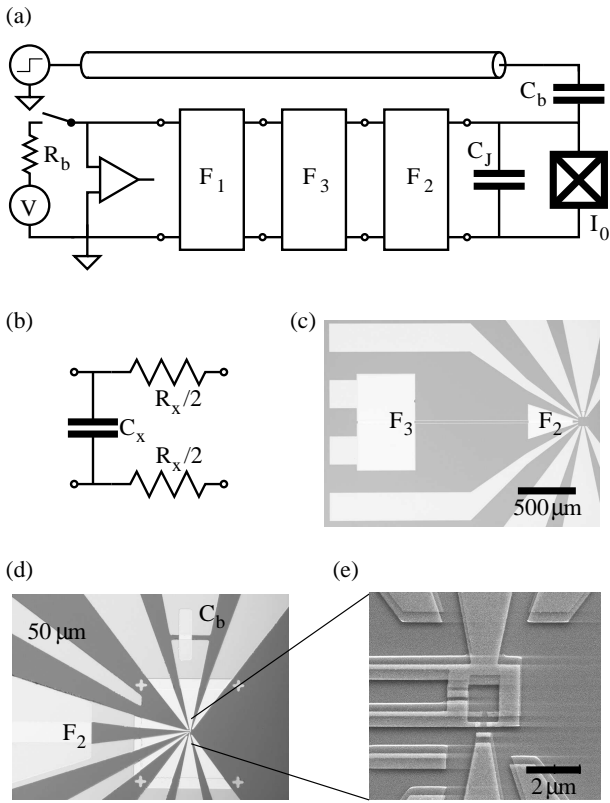}
\caption{(a) Block diagram of the experimental setup. (b) Schematic diagram of the model used to describes the three 
different filters $F_1, F_2, F_3$ which define the damping circuit. (c) Micrograph of the sample showing the two 
on-chip RC filters $F_1$ and $F_2$. (d) Magnified view of the center part of the chip, $F_2$ and the bias capacitor. 
(e) Electron microscope picture of the Quantronium (sample III).}
\label{experimentalSetup}
\end{center}
\end{figure}

Experiments investigating junction current-voltage characteristics (IVC) as well as pulsed switching behavior were 
carried out in a dilution refrigerator with $25$ mK base temperature. A block diagram of the measurement setup is 
shown in fig.~\ref{experimentalSetup}(a). A low noise instrumentation amplifier (Burr-Brown INA110, noise temperature 
$1.3$~K at $10$~kHz) is measuring the voltage across the sample while the sample is biased by a room temperature 
voltage source either via the bias capacitor $C_b$, or  in series with a bias resistor $R_b$.  The capacitor bias 
method was used for experiments with fast current pulses of duration $\tau_p=25$ ns, while the conventional resistor 
bias method was used for long pulse experiments with $\tau_p=20$ \textmu s, as well as for IVC measurements.

Three different samples are discussed in this paper which differ primarily in the range of the measured switching 
current (3 nA to 120 nA), and in the type of circuit used for the damping of the phase dynamics.  These different 
damping circuits are labeled in the order in which they were implemented, and are represented in 
fig.~\ref{experimentalSetup}(a) as the blocks $F_1, F_2, F_3$.  These environments can be modeled as RC filters with 
different cut-off frequencies, as schematically be represented in fig.~\ref{experimentalSetup}(b).

\begin{table*}
\renewcommand{\footnoterule}{}
\renewcommand{\thempfootnote}{\alph{mpfootnote}}
\begin{center}
\begin{tabular}{|c|c|c|c|c|c|c|c|c|c|c|c|c|c|}
\hline
& & \multicolumn{3}{|c|}{CPT} & \multicolumn{3}{|c|}{Shunt JJ} & \multicolumn{2}{|c|}{$\mbox{F}_1$} & 
\multicolumn{2}{|c|}{$\mbox{F}_2$} & \multicolumn{2}{|c|}{$\mbox{F}_3$} \\
\cline{3-14}
Sample & Type & $E_J/E_C$ & $I_0$ & $I_{sw}$ & $E_J/E_C$ & $I_0$ & $I_{sw}$ & $R_1$ & $C_1$ & $R_2$ & $C_2$ & $R_3$ & 
$C_3$\\
\hline
\hline
~I~ & ~CPT~ & ~32.9~ & ~58.5~ & ~3~ & ~-~ & ~-~ & ~-~ & ~60~ & ~1~ & ~-~ & ~-~ & ~-~ & ~-~ \\ 
\hline
~II~ & ~CPT~ & ~29~ & ~51.6~ & ~4.2~ & ~-~ & ~-~ & ~-~ & ~60~ & ~1~ & ~7.2~ & ~0.24~ & ~-~ & ~-~ \\
\hline
~III~ & ~Quantronium~ & ~2.2~ & ~21~ & ~12~ & ~30.3~ & ~158~ & ~120~ & ~1000~ & ~3~ & ~7.2~ & ~0.24~ & ~600~ & ~1.4~ 
\\
\hline
\end{tabular}
\footnote[0]{Currents are in [nA], resistances in [$\Omega$] and capacitances in [nF]}
\caption{An overview over the parameters for the three different samples. Filter $F_1$ resembles the cryostat leads or 
a cold SMD filter. Filter $F_2$ is the on-chip damping circuit and $F_3$ is an on-chip RC-filter.}
\label{table}
\end{center}
\end{table*}

The key parameters for each sample are given in Table \ref{table}.  Sample I consisted of a Cooper pair transistor 
(CPT) embedded in an environment defined solely by the twisted pair leads of the cryostat which is modeled as $F_1$. 
Sample II was a CPT fabricated in parallel with Sample I, having  nearly identical parameters, differing only in that 
sample II was embedded in a micro-fabricated on-chip HF-damping circuit $F_2$.   Sample III is a Quantronium 
\cite{vion:quantronium:02} embedded in the same HF-damping $F_2$ used with sample II, but with an additional 
micro-fabricated on-chip low-pass filter $F_3$.  The on-chip RC-environments $F_2$ and $F_3$ used for samples II and 
III and the bias capacitor were fabricated with a two-step optical lithography process.  The capacitors were actually 
two capacitors in series, formed by a plasma-oxidized $\mbox{Al}$ ground plane covered with a $\mbox{Au}$ top plate.  
The top plates are connected to the rest of the circuit via resistors which are formed from the same $\mbox{Au}$  film 
as the top plate, having a typical sheet resistance of $1.2~\Omega / \Box$.  The capacitors of $F_3$ could be measured 
quite accurately, from which we obtain a specific capacitance of 13.6 fF/\textmu m$^2$ that is used to determine all 
on-chip capacitors.  Figure~\ref{experimentalSetup}(c) shows the essential parts of the chip and the components 
defining the high-frequency environment. The bright rectangular area on the left side is the top plate of the 
capacitor, and the thin leads leading to the right are the resistors of filter $F_3$.  
Figure~\ref{experimentalSetup}(d) shows in detail the biasing capacitor $C_b$ The bright trapezoidal area on the left 
is the top plate of the capacitor $C_2$ and the areas surrounded by dashed lines are damping resistors $R_2/2$.  
Figure~\ref{experimentalSetup}(e) is an electron microscope picture showing the $\mbox{Al/Al}_2\mbox{O}_3\mbox{/Al}$ 
tunnel junctions, which were fabricated in a third layer of electron beam lithography, with the standard two-angle 
evaporation through a shadow mask.  Figure~\ref{experimentalSetup}(e) shows the quantronium circuit of sample III.

\begin{figure}[!hb]
\begin{center}
\includegraphics[width=0.45\textwidth]{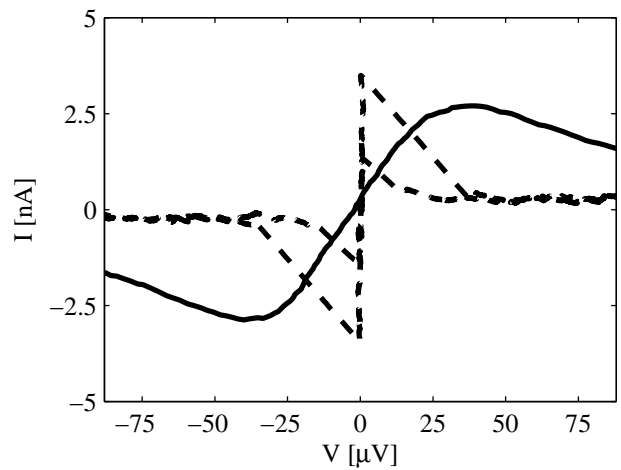}
\caption{IV curves of sample I without designed RC environment (solid line) and of sample II with specially designed 
RC environment (dashed line).}
\label{IVcurves}
\end{center}
\end{figure}

The effect of the on-chip HF damping from $F_2$ on the phase dynamics as can be seen in fig.~\ref{IVcurves} where the 
IVC of samples I (solid line) and sample II (dashed line) are shown.   These two CPT samples differ essentially by the 
presence of $F_2$ in sample II.  We see that the typical phase diffusion shape of the IVC 
\cite{steinbach:supercurrent:01} of sample I is absent in sample II which shows a sharp supercurrent and hysteretic 
switching. The presence of the on-chip environment in sample II effectively reduces phase diffusion as can be 
explained by a phase-space topology as shown in figure~\ref{phaseSpace1}(d). However, the very low value of $I_{sw}$ 
is a direct indication of excessive noise in the circuit. Therefore the on-chip low-pass filter $F_3$ was implemented 
in sample III, improving the switching current to a value of $75\%$ of the critical current. In the remainder of this 
paper we concentrate on investigating the switching behavior of sample III.

The ability to suppress phase diffusion opens up the possibility to study fast switching with HF-overdamped phase 
dynamics for the first time.  We used the pulse and hold method to measure switching probabilities of sample III as a 
function of the amplitude of the switch pulse for two cases:  A long pulse of 20 \textmu s where the switching was 
found to be controlled by equilibrium thermal escape, and a short pulse of 25ns, where the switching is clearly a 
non-equilibrium process.  

\begin{figure}
\begin{center}
\includegraphics[width=0.47\textwidth]{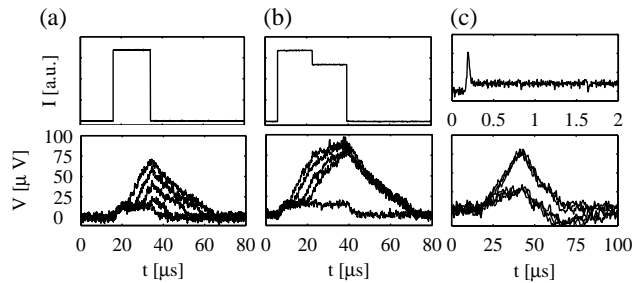}
\caption{(a) Square pulse and response. (b) Switch pulse with hold level and response. (c) Switch pulse with hold 
level and response, generated by the capacitive bias method.  The finite voltage in the non-switch case is due to the 
series filter resistance and the two-point measurement setup.}
\label{pulseComparison}
\end{center}
\end{figure}

The long pulses were formed by applying a square voltage pulse through the bias resistor.  The response to a simple 
square pulse is shown in fig.~\ref{pulseComparison}(a), where the applied voltage pulse is shown, and several scope 
traces of the measured voltage over the CPT are overlayed.  Here we see that the switching causes an increase in the 
voltage over the sample which can occur at any time during the applied pulse.  In order to do statistics we want to 
unambiguously count all switching events. Late switching events are difficult to distinguish from non-switching events 
as the voltage does not have time to rise above the noise level.  We can add a trailing hold level as shown in 
fig.~\ref{pulseComparison}(b).  This hold level and duration must be chosen so that there is zero probability of 
switching on the hold part of the pulse.  The response to such a pulse shows that it is now easy to distinguish switch 
from non-switch events.  In this case the hold level is used simply to quantize the output, and the switching which 
occurs during the initial switch pulse is found to be a thermal equilibrium escape process as discussed below. 

The fast pulses were formed with a new technique where a voltage waveform consisting of a sharp step followed by 
linear voltage rise is programmed in to an arbitrary waveform generator.  The slope on the sharp step 
$(dV/dt)_{pulse}$ is typically 6--7 times larger than the linear rise during the hold, $(dV/dt)_{hold}$.  The voltage 
waveform is propagated to the chip through a coax cable having negligible dispersion for the sharp 25 ns voltage step 
used.  An on-chip bias capacitor $C_b$ will differentiate the voltage waveform to give a sharp current pulse followed 
by a hold level, $I=C_b dV/dt$, which is shown in fig.~\ref{pulseComparison}(c). From the measured step amplitude 
needed to switch the junction and the value of $C_b=1.4$~pF, we calculate a pulse amplitude of 360~nA through $C_b$. 
Due to the symmetry of the filter stages $F_1$ to $F_3$, only half of this 25~ns pulse current flows through the 
junction, with the other half flowing through the filter to ground. Thus the peak current through the junction during 
the 25~ns pulse $I_p = 180$~nA, which is larger than $I_0$.  Exceeding $I_0$ for this very short time is not 
unreasonable, bearing in mind that the circuit is heavily overdamped at high frequencies, and a strong kick will be 
needed to overcome damping and bring the phase particle close to the saddle point.

The hold level for these fast pulses is $40$~\textmu s,very much longer than the switch pulse, and its duration is set 
by the time needed for the response voltage to rise above the noise level.  The rate of this voltage rise depends on 
the hold current level because after the switch we are essentially charging up the second stage filter and leads, 
$F_3$ and $F_1$, with the hold current, $I_{hold}=  C_b (dV/dt)_{hold} = 56$~nA.  For the low level of hold current 
used in these experiments, we can follow the voltage rise at the junction with the 100 kHz bandwidth low noise 
amplifier at the top of the cryostat.  Typically we turn off the hold current and reset the detector when the sample 
voltage is ~30 \textmu V, so that the junction voltage is always well below the gap voltage $V_{2 \Delta}=400$ \textmu 
V, and therefore quasi-particle dissipation during the hold can be neglected. 

\begin{figure}
\begin{center}
\includegraphics[width=0.45\textwidth]{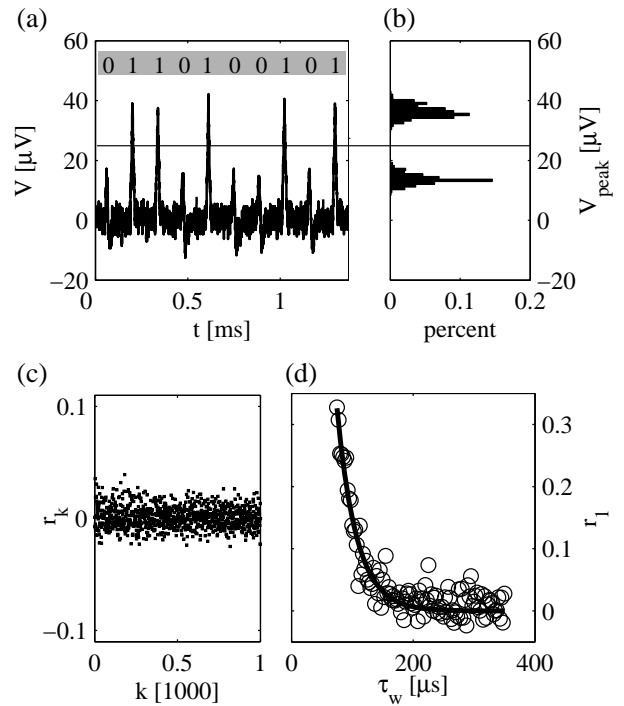}
\caption{(a) Response of the sample to a pulse sequence resulting in switches (1) and no-switches (0) of the sample. 
(b) Peak voltage obtained during a current pulse, indicating good separation between the switch and non-switch signal. 
(c) Auto-correlation function $r_k$. (d) Correlation coefficient $r_1$ vs. the wait time, with fitted exponential 
decay function.}
\label{measurement}
\end{center}
\end{figure}

Pulsed switching measurements were performed were a sequence of $10^3$ to $10^4$ identical pulse-hold-reset cycles was 
applied to the sample while recording the voltage response of the sample.  A threshold level was used to distinguish 
switching events (1) from non-switching events (0) as depicted in fig.~\ref{measurement}(a).  The maximum response 
voltage achieved during each cycle is found and a histogram of these values is plotted as seen in 
fig.~\ref{measurement}(b).  The hold level and duration are adjusted so as to achieve a bimodal distribution in the 
histogram, with zero events near the threshold level, meaning that there is zero ambiguity in determining a switch 
event from a non-switch event.  We further check that the hold level itself, without the leading switch pulse, gives 
no switches of the sample.  The sequence of switching events is stored as a binary sequence $Y_i$ in temporal order.  
From this sequence we can calculate the switching probability, 
\begin{equation}
P={1 \over N}\sum\limits_{i=1}^N {Y_i}
\end{equation}
and the auto-correlation coefficients,
\begin{equation}
r_k=\frac{\sum_{i=1}^{N-k}(Y_i-\overline{Y})(Y_{i+k}-\overline{Y})}{\sum_{i=1}^N(Y_i-\overline{Y})^2}.
\end{equation}
where $k$ is the "lag" between pulses. The auto-correlation is a particularly important check for statistical 
independence of each switching event.  A plot of $r_k$ for $k=1...1000$ is shown in figure~\ref{measurement}(c) and 
the randomness and low level of $r_k$ indicates that all switching events are not influenced by any external periodic 
signal.  When the circuit is not working properly, pick up of spurious signals up to the repetition frequency of the 
measurement, clearly shows up as a periodic modulation in the auto-correlation $r_k$. Of particular importance is the 
correlation coefficient for lag one $r_1$ which tells how neighboring switching events influence one another.  
Fig.~\ref{measurement}(d) shows $r_1$ as a function of the wait time $\tau_w$ between the end of the hold level and 
the start of the next switch pulse. For large values of $\tau_w$, $r_1$ fluctuates around 0 not exceeding ~0.05, which 
shows that any influence of a switching or non-switching event on the following measurement, is statistically 
insignificant.  As $\tau_w$ is decreased however, a positive correlation is observed, with $r_1$ increasing 
exponentially with shorter $\tau_w$.  Positive correlation indicates that a switching event (a "1") is more likely to 
be followed by another switching event.  Fig.~\ref{measurement}(d) shows a fit to correlation $r_1$ to the function
\begin{equation}
r_1=3.345\cdot \exp \left(-\frac{\tau_w}{33.3\mbox{\textmu s}}\right).
\end{equation}
We can extrapolate the fit to the time $\tau^*=40.25$~\textmu s where the auto-correlation becomes $r_1=1$, meaning 
that once the circuit switches it will always stay in the 1-state.  In our experience, increasing the capacitance of 
filter $F_1$ causes $\tau^*$ to increase, from which we infer that the increase in the correlation $r_1$ for short 
$\tau_w$ is resulting from errors where the detector is not properly reset because it does not have time to discharge 
the environment capacitance before a new pulse is applied.  For the experiment shown in figure \ref{measurement} the 
time constant of the environment was estimated to be 3~\textmu s.  These observations indicate that it is necessary to 
bring the junction voltage very close to zero before the retrapping will occur, and the detector will reset.  For good 
statistics many pulses are required and a short duty cycle is desirable in order to avoid effects from low frequency 
noise as discussed section III.  By studying the correlation coefficient $r_1$ in this way, we can choose an optimal 
duty cycle.

\section{Analysis}\label{analysis}

\begin{figure}
\begin{center}
\includegraphics[width=0.45\textwidth]{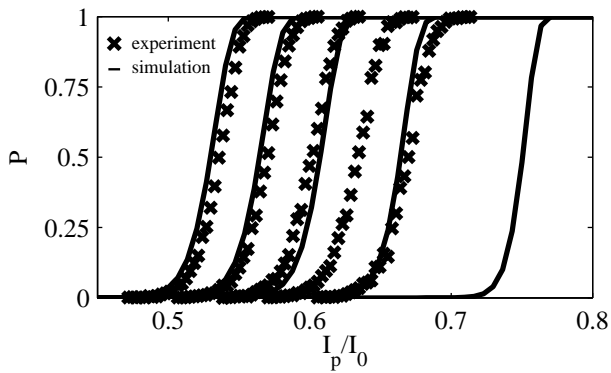}
\caption{Switching probability as a function of pulse height in the temperature range $T=100,~200,~300,~400,~500$~mK 
(right to left) for a pulse duration $\tau_p=20$ \textmu s. Crosses show measured data and simulated data is shown as 
solid lines.}
\label{sCurves}
\end{center}
\end{figure}

The switching probabilities were thus measured and the dependence on the amplitude of the switch pulse, $P(I_p)$ was 
studied as as a function of temperature.  Each  measurement of $P(I_p)$ began with a pulse sequences having pulse 
amplitude resulting in a switching probability $P=0$, and the pulse amplitude was successively increased until $P=1$.  
The measurement produces an "S-curve" as shown in figure~\ref{sCurves}, where the experimental data for the long pulse 
duration $\tau_p=20$~\textmu s is shown with crosses.  The S-curves were taken at temperatures 100, 200, 300, 400 and 
500~mK (right to left) respectively.

We compare the measured data to theoretical predictions based on thermal escape as discussed in section IV.  The 
filter $F_1$ causes a rounding of the applied square voltage pulse, which is accounted for by calculating the escape 
probability for a time dependent current \cite{gotz:phd:97},
\begin{equation}
P=1-\exp{\left( -\frac{1}{di/dt}\int_0^i \Gamma(i')di'\right)}
\end{equation}
where the escape rate $\Gamma$ can be found using either eqns.~\ref{Kramers} or \ref{overdampedEscape}.   The 
simulated S-curves using eqn. \ref{overdampedEscape} are plotted in figure~\ref{sCurves} as solid lines for the 
temperatures corresponding to the measured data.  Sample parameters used for this calculation are the measured bias 
(including filter) resistance $R_1=11600~\Omega$, the measured high frequency damping resistor, $R_2=7.2~\Omega$, the 
high frequency damping capacitance $C_2=0.207$~nF, the junction capacitance $C_J=30$~fF, and the calculated critical 
current $I_0=148$~nA.  The critical current $I_0=148$~nA is not the bare critical current $I_0=158$~nA  since the 
quantronium was biased at a magnetic field such that a persistent current of $\approx 10$~nA was flowing in the loop.  
These parameters are all independently determined, and not adjusted to improve the fit.   However, the capacitance of 
filter $F_1$ was uncertain, having a nominal value of 10nF, and unknown temperature dependence below $4$~K.  Cooling 
the same capacitance to $4$~K, we observed a decrease of $C_1$ by around $10\%$. This capacitor $C_1$ determines the 
rounding of the square voltage pulse, and thus the time dependence of the current applied to the junction.   We found 
that it was necessary to assume $C_1=3$~nF  in order for the simulations to agree with experiment.  This low value of 
$C_1$ at low temperatures is not unreasonable, as circuit simulations with the nominal value of 10 nF showed that the 
initial pulse would not exceed the hold level, which clearly is not possible because excellent latching of the circuit 
was observed.  

\begin{figure}
\begin{center}
\includegraphics[width=0.45\textwidth]{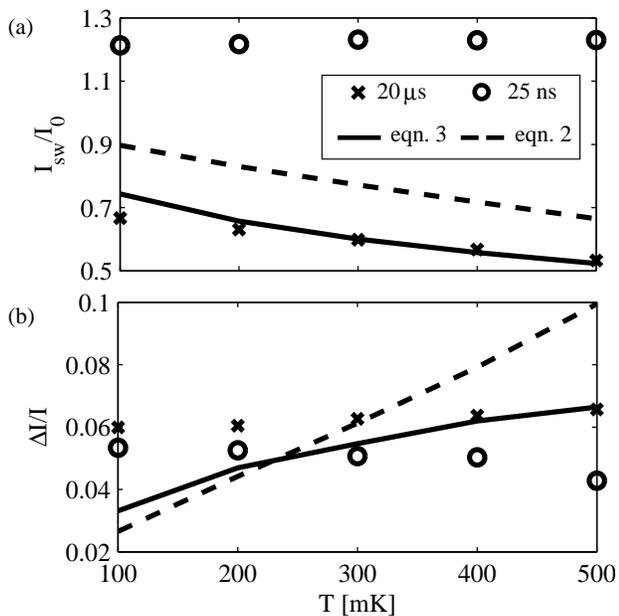}
\caption{Switching current normalized to critical current (a) and relative resolution (b) of sample III. Crosses 
indicate measured values for $\tau_p=20$~\textmu s, solid and dashed lines are calculated values using a 
generalization of Kramers' large friction result and Kramers' original  result, respectively. Circles are measured 
values for $\tau_p=25$~ns.}
\label{IoverI0}
\end{center}
\end{figure}

From the experimental and simulated S-curves, we define the switching current of the sample as the pulse amplitude 
that gives 50 \% switching $P(I_{sw})=0.5$ and the resolution is defined from the S-curve by $\Delta 
I=I_p(P=0.9)-I_p(P=0.1)$.
A comparison of experimental and theoretical $I_{sw}$ vs. $T$ and $\Delta I/I_{sw}$ vs. $T$ is shown in 
fig.~\ref{IoverI0}. We see
that the experimental data for the long pulses (points marked by an X) are in reasonably good agreement with the 
simulated values when the theory of switching in an environment with frequency dependent damping  is used (escape from 
a meta-potential, equation \ref{overdampedEscape}) which is plotted as a solid line in fig.~\ref{IoverI0}(a).  We note 
that for the 20~\textmu s pulses, escape occurs at bias currents $i\approx 0.7$, where the phase space has a topology 
as shown in figure \ref{phaseSpace1}(d).  Hence we can neglect phase diffusion and escape is from a saddle point, so 
that the non-absorbing boundary condition assumed in the theory is valid.  For comparison, we use the overdamped 
Kramers formula (equation \ref{Kramers}) to simulate the S-curve and calculate $I_{sw}$ and $\Delta I$, which is shown 
by the dashed line in fig.~\ref{IoverI0}.  Here the prefactor $\kappa (Q)$ is given in ref.\cite{kautz:noise:90} and 
we have used the high frequency quality factor $Q_1=0.027$ as determined by the resistor $R_2$ only.  We see that the 
Kramers formula overestimates $I_{sw}$ by some 25\% (fig.~\ref{IoverI0}(a)) and is worse than the simulation based on 
eqn.~\ref{overdampedEscape}, in reproducing the temperature dependence of $\Delta I$ (fig.~\ref{IoverI0}(b)).  In 
fact, the experimental data for the $20$~\textmu s pulses only shows a weak increase in $\Delta I$ over the 
temperature range studied, whereas both theoretical curves predict a slight increase in $\Delta I$.  Thus an 
equilibrium thermal escape model explains the data for long, 20 \textmu s pulses reasonably well and  the data is 
better explained by the theory of escape with frequency dependent damping, than by the simpler theory embodied in the 
overdamped Kramers formula.  However the correspondence with the former theory is not perfect.   We may explain these 
deviations as being due to the fact that the quality factor $\alpha=4.49$ (see section~\ref{fluctuations}) does not 
really satisfy the condition for validity of the theory, $\alpha \gg 1$.

Experimental data for the short pulses of duration $\tau_p=25$~ns generated by the capacitive bias method is plotted 
in fig.~\ref{IoverI0} as circles. Here we see that the value of $I_{sw}$ is constant in the temperature range studied, 
indicating that escape is not from a thermal equilibrium state. For the ideal phase space topology, as shown in figure 
\ref{phaseSpace1}(d), the initial pulse would bring the phase particle arbitrarily close to the saddle point S for the 
hold bias level.  If the separation in to the basins of attraction occurs before thermal equilibrium can be 
established, we would not expect temperature dependence of $I_{sw}$. In this case, the width of the switching 
distribution will be determined not by thermal fluctuations, but by other sources of noise, such as random variation 
in the height of the switch pulse.  These variations are significant because the 1/f noise from the waveform generator 
must be taken in to account when generating the train of pulses over the time window of the measurement which was 
about 0.5 sec.  In our experiments however, we may not have achieved a constant hold level since the voltage ramp from 
the waveform generator is not perfectly smooth.  Knowing the bias capacitor we can calculate an average hold level of 
$i_{hold}=0.35$, somewhat lower than the critical value of $i_{hold}=0.67$ necessary to achieve the phase space 
topology of figure \ref{phaseSpace1}(d).  Nevertheless, we observe excellent latching of the circuit for these 25~ns 
switch pulses.  We conclude that the observed temperature independence of $I_{sw}$, and the fact that $I_{sw}$ exceeds 
$I_0$ by 20\%  is consistent with a very rapid switching of the junction. 

We can rule out excessive thermal noise as a reason for the temperature independent value of $I_{sw}$ for the short 
pulses.  By measuring the gate voltage dependence of $I_{sw}$ as a function of the temperature, a clear transition 
from $2e$ to $e$ periodicity was observed in the temperature range 250 mK to 300 mK. For the size of the 
superconducting island used in this experiment, we can estimate a crossover temperature $T^* \approx 300$~ mK, above 
which the free energy difference between even and odd parity goes to zero.\cite{tuominen:parity:92} Hence, we know 
that the sample is in equilibrium with the thermometer below $T^*$, and therefore heating effects that might occur in 
the short pulse experiments, can not explain the fact that the observed $I_{sw}$ is independent of temperature.

Thus we have achieved a very rapid, 25 ns measurement time of the switching current, which should be sufficient for 
measurement of the quantum state of a quantronium circuit.  For qubit readout, not only the measurement time is 
important, but also the resolution of the detector.  For the 25 ns pulse, we obtained the resolution of $\Delta 
I/I_{sw}=0.055$, or $\Delta I=9.9$ nA.  This implies that single shot readout is possible for a Quantronium with 
parameters $E_C=0.5$~K and $E_J/E_C=2.5$ where the switching current of the two qubit states at the optimal readout 
point differ by 9.6~nA. Numerous experiments were made with microwave pulses and continuous microwave radiation to try 
and find the qubit resonance. However, due in part to uncertainty in the qubit circuit parameters (level separation) 
and in part to jumps in background charge, no qubit resonance was detected in these experiments.

\section{Conclusion}
Fast and sensitive measurement of the switching current can be achieved with a pulse-and-hold measurement method, 
where an initial switch pulse brings the JJ circuit close to an unstable point in the phase space of the circuit 
biased at the hold level.  This technique exploits the infinite sensitivity of a non-linear dynamical system at a 
point of bifurcation, a common theme in many successful JJ qubit detectors.  We have shown that with properly designed 
frequency dependent damping, fast switching can be achieved even when the high frequency dynamics of the JJ circuit 
are overdamped.   With an on-chip RC damping circuit, we have experimentally studied the thermal escape process in 
overdamped JJs.  A capacitor bias method was used to create very rapid 25ns switch pulses. We demonstrated fast 
switching in such overdamped JJs for the first time, where the switching was not described by thermal equilibrium 
escape. The methods presented here are a simple and inexpensive way to perform sensitive switching current 
measurements in Josephson junction circuits. While we have shown that the sensitivity can be high, the effect of 
back-action of such a detector is still unclear and might be a reason why no quantum effects were observed. In 
contrast to the readout strategy presented here, all other working qubit-readout strategies, both static switching and 
dispersive, are based on underdamped phase dynamics.

\begin{acknowledgments}
This work has been partially supported by the EU project SQUBIT II, and the SSF NanoDev Center. Fabrication and 
measurement equipment was purchased with the generous support of the K. A. Wallenberg foundation. We acknowledge 
helpful discussion with D. Vion and M. H. Devoret and theoretical support from H. Hansson and A. Karlhede.
\end{acknowledgments}

\end{document}